\documentclass[twocolumn,prl,notitlepage,amsfonts,citeautoscript,superscriptaddress,showpacs]{revtex4-1}
\usepackage{natbib}

\usepackage{listings}
\usepackage{color}

\definecolor{dkgreen}{rgb}{0,0.6,0}
\definecolor{gray}{rgb}{0.5,0.5,0.5}
\definecolor{mauve}{rgb}{0.58,0,0.82}

\usepackage{amsmath,amssymb,mathrsfs}
\usepackage{latexsym}
\usepackage{graphicx} 
\usepackage{epstopdf}
\usepackage{graphicx,epstopdf,color}
\usepackage{amsfonts}
\usepackage{amsmath,amssymb,mathrsfs}
\usepackage[breaklinks]{hyperref}

\usepackage{color}
\usepackage{url}
\usepackage{bm}

\begin{document}

\title{Fluxon-Based Quantum Simulation in Circuit QED}

\author{Alexandru Petrescu}  
\email{tpetresc@princeton.edu}
\affiliation{Department of Electrical Engineering, Princeton University, 08544 Princeton, New Jersey, USA}
\author{Hakan~E.~T\"ureci}
\affiliation{Department of Electrical Engineering, Princeton University, 08544 Princeton, New Jersey, USA}
\author{Alexey V. Ustinov}
\affiliation{Physikalisches Institut, Karlsruhe Institute of Technology, 76131 Karlsruhe, Germany}
\affiliation{Russian Quantum Center, National University of Science and Technology MISIS, 119049 Moscow, Russia}
\author{Ioan~M.~Pop}
\affiliation{Physikalisches Institut, Karlsruhe Institute of Technology, 76131 Karlsruhe, Germany}

\date{\today}
\begin{abstract}
Long-lived fluxon excitations can be trapped inside a superinductor ring, which is divided into an array of loops by a periodic sequence of Josephson junctions in the quantum regime, thereby allowing fluxons to tunnel between neighboring sites. By tuning the Josephson couplings, and implicitly the fluxon tunneling probability amplitudes, a wide class of 1D tight-binding lattice models may be implemented and populated with a stable number of fluxons. We illustrate the use of this quantum simulation platform by discussing the Su-Schrieffer-Heeger model in the 1-fluxon subspace, which hosts a symmetry protected topological phase with fractionally charged bound states at the edges. This pair of localized edge states could be used to implement a superconducting qubit increasingly decoupled from decoherence mechanisms.
\end{abstract}
\maketitle

With recent advances in state preparation and measurement techniques,  circuit quantum electrodynamics (cQED) architectures \cite{blais_et_al_2004,blais_et_al_2007} are becoming increasingly attractive for quantum information processing and quantum simulation \cite{you_nori_2011}. Other platforms for quantum simulation  include ultracold atoms in traps and optical lattices \cite{bloch_et_al_2012}, trapped ions \cite{johanning_et_al_2009, blatt_roos_2012}, Josephson junction arrays \cite{fazio_van_der_zant_2001}, or photonic systems \cite{georgescu_et_al_2014}. One of the main efforts in quantum simulation has been the implementation of interacting, strongly-correlated models, which possess rich physics, but are in general analytically intractable. 

There is an increasing list of proposals based on the cQED architecture, which notably includes analogues of the seminal boson Hubbard model \cite{fisher_et_al_1989} for the superfluid to insulator transition of lattice bosons with repulsive contact interactions \cite{hartmann_et_al_2006,angelakis_et_al_2007,hartmann_et_al_2008,makin_et_al_2008,koch_le_hur_2009,houck_et_al_2012,schiro_et_al_2012,schmidt_koch_2013}, the fermion Hubbard model \cite{reiner_et_al_2016}, or topological order \cite{cho_et_al_2008, hayward_et_al_2012}. Recently, several implementations have successfully shown proof-of-concept quantum simulation of dissipative phase transitions 
 \cite{fitzpatrick_et_al_2017}, molecules \cite{omalley_et_al_2016} or fermionic tight-binding models \cite{barends_et_al_2015}, and the Rabi model in the strong and ultrastrong coupling regimes \cite{niemczyk_et_al_2010, yoshihara_et_al_2017, forn-diaz_et_al_2017, braumueller_et_al_2016,langford_et_al_2016}, heralding studies of spin-boson and Kondo physics \cite{le_hur_et_al_2016}. 

Microwave photons, the physical building block for cQED quantum Hamiltonians, are nevertheless subjected to intrinsic dissipation. 
One solution to circumvent the limitations imposed by photon loss is to stabilize quantum states using bath-engineering schemes for single qubits \cite{murch_et_al_2012, leghtas_et_al_2015}, or qubit arrays \cite{kimchi-schwartz_et_al_2016,aron_et_al_2014,aron_et_al_2016}.

\begin{figure}[t!]
  \includegraphics[width=0.90\linewidth]{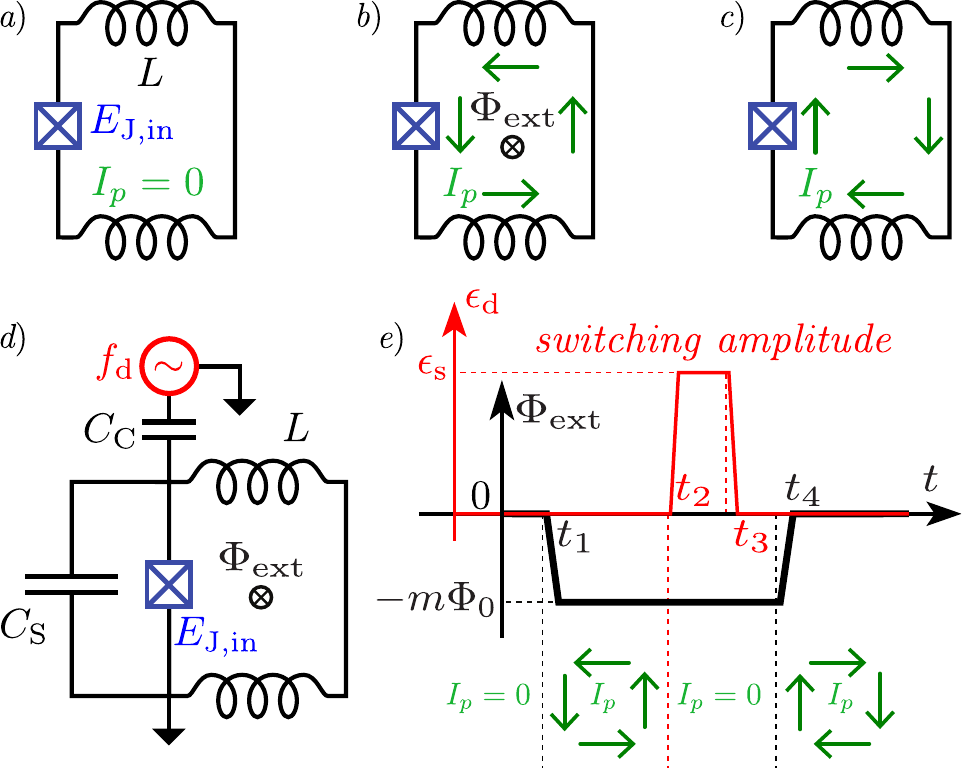}
  \caption{Fluxon state preparation. The top row shows \label{Fig:1} the three possible configurations of the superinductor loop: \textit{a}) no persistent current $I_p = 0$, \textit{b}) persistent current under external flux, and \textit{c}) $m$ fluxons trapped inside of the loop with zero external field. \textit{d}) Circuit setup for insertion of fluxons under external flux and drive. \textit{e}) Protocol for drive amplitude $\epsilon_d(t)$ (red) and external flux $\Phi_\text{ext}(t)$ (black) for the insertion of $m$ fluxons. The drive frequency $f_\text{d} \approx \frac{1}{2\pi} (L_\text{J,in} C_{\text{S}})^{-1/2}$ corresponds to the eigenmode of the radiofrequency (RF) resonator consisting of the input junction $E_\text{J,in}$ and the shunt capacitance $C_\text{S}$. The corresponding persistent currents are represented by the green arrows [Supplemental Material].}
\end{figure}

In this Letter, we propose an alternative way to simulate lattice models, where the ground state of the effective Hamiltonian is unaffected by photon losses. Specifically, we show how to engineer arbitrary one-dimensional tight-binding models for quantum fluxons, \textit{i.e.} $2\pi$-kinks in the superconducting phase order parameter. Fluxons correspond to remarkably stable quantized persistent currents $I_p$ flowing around superconducting loops containing Josephson junctions [Fig.~1a-c)]. In order to load a certain number of fluxons $m$ inside the ring, one can use a protocol very similar to the one demonstrated in Ref.~\cite{masluk_et_al_2012} for the reset of a superinductor loop to its ground state with $m=0$ [Fig.~\ref{Fig:1}d)-e)]. We expect this protocol to successfully implement the desired $m$-fluxon state with a probability in excess of 90\%, stable for an extended duration of time, on the order of hours or even days \cite{masluk_et_al_2012}. 

In the classical regime, fluxons constitute the basis for rapid single flux quantum electronics (RSFQ) \cite{likharev_semenov_1991}, where current biases close to the critical current prompt fluxon mobility. Although classical fluxon dynamics is inherently dissipative, the associated heating is low enough to make them attractive for state-of-the-art classical information processing \cite{mukhanov_2011}.

Quantum fluxons are significantly more fragile. Following their first implementation a decade ago \cite{wallraff_et_al_2003},  their use in devices has remained limited, with few exceptions, notably in the recent design of a qubit readout circuit \cite{fedorov_et_al_2014}. One of the main challenges in the development of quantum fluxon electronics was the absence of reliable superinductors, inductors $L$ with an RF impedance comparable to the resistance quantum: $L\omega \geq R_Q = h/(2e)^2 \approx 6.5$~k$\Omega$. The remarkable recent progress in superinductor design and fabrication \cite{manucharyan_et_al_2009,masluk_et_al_2012,bell_et_al_2012}, including their use in artificial crystals and molecules \cite{meier_et_al_2015, kou_et_al_2017}, renders possible the physical implementation of the quantum fluxon platform proposed in this Letter.

\begin{figure}[t!]
  \includegraphics[width=\linewidth]{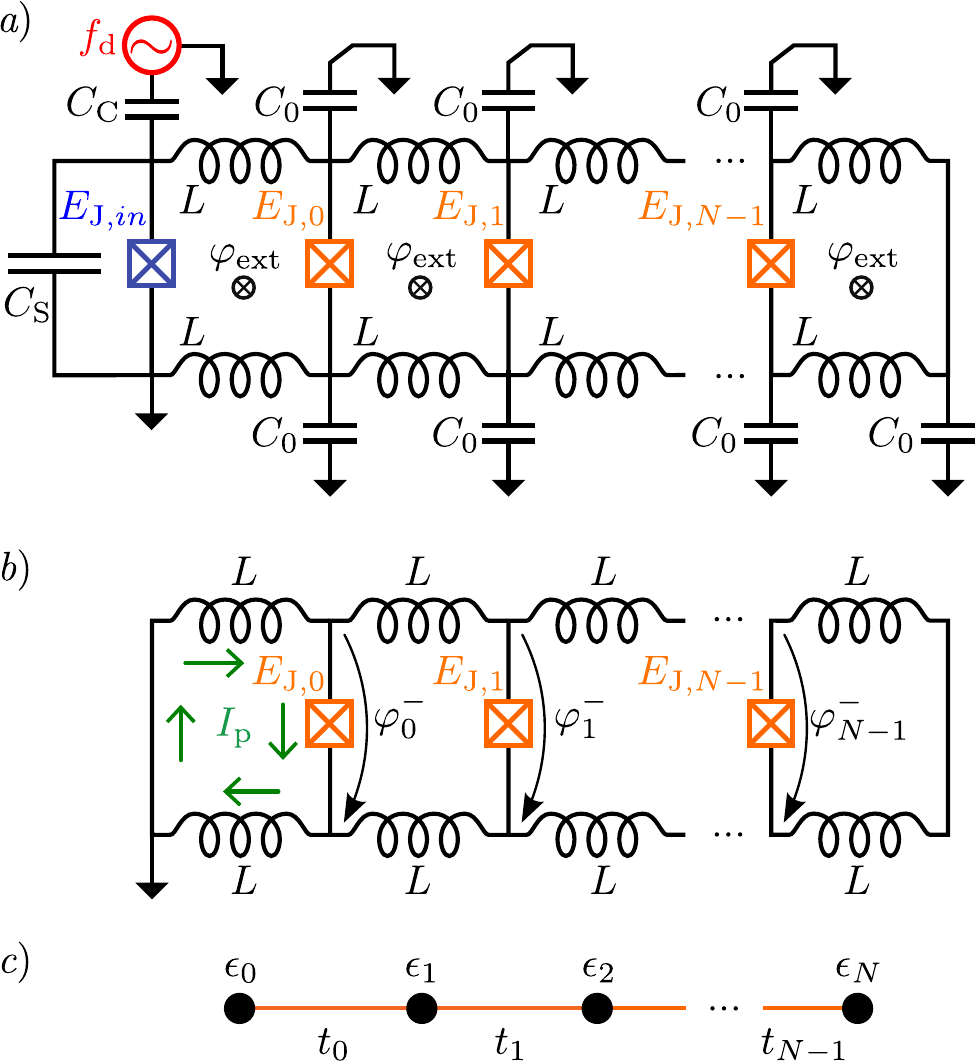}
    \caption{Superconducting circuit implementation of an effective tight-binding model for fluxons. \textit{a}) The circuit of Figure~\ref{Fig:1} generalizes to a superinductor ring encompassing loops separated by Josephson junctions. The fluxon ``input'' junction is shown in blue, the ``lattice'' junctions $E_{\text{J},i}$ are depicted in orange. \textit{b}) Circuit representation of the simplified model of Eq.~(\ref{Eq:HMinus}); the branch fluxes $\varphi_{i}^-$ are the degrees of freedom describing fluxon dynamics through the one-dimensional array. \textit{c}) The equivalent tight-binding model for fluxons, where every site corresponds to a loop in \textit{b}); the on-site and tunneling energy scales are the ones appearing in Eq.~(\ref{Eq:TBH}).\label{Fig:2}}
\end{figure}

The key insight of our proposal is to implement a tight-binding model for long-lived quantum fluxons trapped inside a superinductor ring. The ring is divided into smaller loops by a periodic sequence of quantum Josephson junctions (see Fig.~\ref{Fig:2}), with $E_{\text{J},i} / E^-_{\text{C},i} \lesssim 10 $, where $E_{\text{J},i}$ is the Josephson coupling of the $i^\textit{th}$ junction and $E^-_{\text{C},i} = e^2/[2(C_{\text{J},i}+C_0/2)]$ is the corresponding charging energy.  This allows fluxons to tunnel between neighboring loops, with a tunneling amplitude whose spatial dependence is modulated by the Josephson couplings, which can either be predefined by fabrication, or tuned \textit{in situ} using locally flux-biased SQUID loops. Using this platform, a wide class of 1D tight-binding lattice models could be implemented and populated with a stable number $m$ of fluxons. Additionally, local fast-flux lines would enable the use of the same platform for quantum annealing \cite{santoro_tosatti_2006}.

We now consider a simplified version of the circuit in Fig.~\ref{Fig:2}\textit{a}) in which the antenna and input junction used in the fluxon insertion protocol can be neglected ($L_{\text{J},\text{in}} \ll L$). The quantum Hamiltonian for this circuit follows from a standard quantization procedure \cite{devoret_1997} [Supplemental Material]. The circuit consists of $2N$ superconducting islands denoted by indices $\alpha,i$, with $i=1,...,N$ the longitudinal coordinate and $\alpha=1,2$ the transverse coordinate. The degrees of freedom are canonically conjugate pairs of superconducting phase and Cooper pair number operators on the superconducting islands, obeying  $[\varphi_{\alpha,i}, n_{\beta,j}] = i \delta_{\alpha \beta} \delta_{ij}$.
We introduce linear combinations corresponding to longitudinal and transverse modes, respectively:
\begin{eqnarray}
  \label{Eq:Rot}
  \varphi^{\pm}_{i} = \varphi_{1,i} \pm \varphi_{2,i} , \; n^{\pm}_{i} = \frac{n_{1,i} \pm n_{2,i}}{2},
\end{eqnarray}
for which $[\varphi^{\eta}_i, n^{\eta'}_j] = i \delta_{\eta\eta'}\delta_{ij}$ for $\eta,\eta'=\pm$.  The transverse variables $\varphi^{-}_{i}$ and $n^{-}_{i}$ denote the 
branch flux, in units of the superconducting flux quantum $\Phi_0/(2\pi)$, and Cooper pair number difference across the $i^\textit{th}$ Josephson junction, respectively. 

Using the notation introduced in Eq.~(\ref{Eq:Rot}), the circuit Hamiltonian separates as $\mathcal{H}   = \mathcal{H}^+ + \mathcal{H}^-$. The desired effective quantum Hamiltonian is $\mathcal{H}^-$, while $\mathcal{H}^+$ describes the longitudinal ``parasitic'' modes of the transmission line in Fig.~\ref{Fig:2}\textit{b}):
\begin{eqnarray}
  \mathcal{H}^+ 
= \sum_{i=0}^{N-1} 4 E_{\text{C}}^+ (n_i^+)^2 + \sum_{i=0}^{N-2} \frac{E_{\text{L}}}{2} (\varphi_{i+1}^+ - \varphi_{i}^{+} - \varphi_{\text{ext},i+1})^2. \nonumber \\
\end{eqnarray}
$E_{\text{C}}^+ = e^2/C_0$ are Coulomb charging energies, with $C_0$ the capacitance to ground of each superconducting island. $E_\text{L} = [\Phi_0/(2\pi)]^2 / (2 L)$ are inductive energies, and $\varphi_{\text{ext},i} = 2\pi \Phi_{\text{ext},i} / \Phi_0 $ is the external flux. Typical values for the capacitance to ground are $C_0 \sim 10\,\text{aF}$ and for the linear inductance $L \sim 100 \,\text{nH}$ \cite{masluk_et_al_2012}. Since there are $N$ pairs of superconducting islands, the plasma frequency characterizing the excitations of the transmission line scales as
\begin{equation}
  \label{Eq:omp}
  \omega^+ = \frac{1}{N \sqrt{L C_0}} \sim \frac{10^{2}}{N} \,\text{GHz}.
\end{equation} 
The maximum feasible circuit length $N$ results from the necessity to isolate the longitudinal modes from the dynamics in the transverse sector. The typical energy scale in the spectrum of the Hamiltonian of the antisymmetric sector, $\mathcal{H}^-$, is set by the Josephson plasma frequency corresponding to one of the junctions in the array \cite{devoret_et_al_1984} 
$\omega^- \sim 10\, \text{GHz}$. We therefore require $\omega^+ \gtrsim \omega^-$, implying a conservative constraint $N \lesssim 10$.

Secondly, $\mathcal{H}^-$ is the Hamiltonian describing the phase difference across the Josephson junctions [see Fig.~\ref{Fig:2}b) for convention], which we express as \begin{equation}
\label{Eq:HMinus}
\mathcal{H}^- = \mathcal{T}^- + \mathcal{V}^-,
\end{equation} 
whose terms are:
\begin{eqnarray}
  \mathcal{T}^- 
  = \sum_{i=0}^{N-1} 4 E^-_{\text{C},i} (n_i^-)^2,
\end{eqnarray}
with $E^-_{\text{C},i} = e^2 / [2(C_{\text{J},i}+C_{0}/2)] \simeq e^2 / 2 C_{\text{J},i}$ the Coulomb charging energy between the two superconducting islands, and
  \begin{eqnarray}
      \mathcal{V}^- &=& \frac{E_\text{L}}{2} ( \varphi_{0}^{-} - \varphi_{\text{ext},0} )^2 + \frac{E_\text{L}}{2} (\varphi_{N-1}^{-} + \varphi_{\text{ext},N})^2 \;\;\;\;\;\;\;\;\;\;\;\;\;
  \nonumber \\
&&   + \sum_{i=0}^{N-2} \frac{E_\text{L}}{2} ( \varphi_{i+1}^{-} - \varphi_{i}^{-} - \varphi_{\text{ext},i+1})^2  \nonumber \\ && + \sum_{i=0}^{N-1} (1-E_{\text{J},i}) \cos \left( \varphi_i^- \right),
  \label{Eq:Vminus}
  \end{eqnarray}
the potential energy from the inductive and Josephson elements. In the above, $E_{\text{J},i}$ is the Josephson energy of the $i^\textit{th}$ junction which sets the scale of the sine-Gordon nonlinearity. 

\begin{figure}[t!]
  \includegraphics[width=\linewidth]{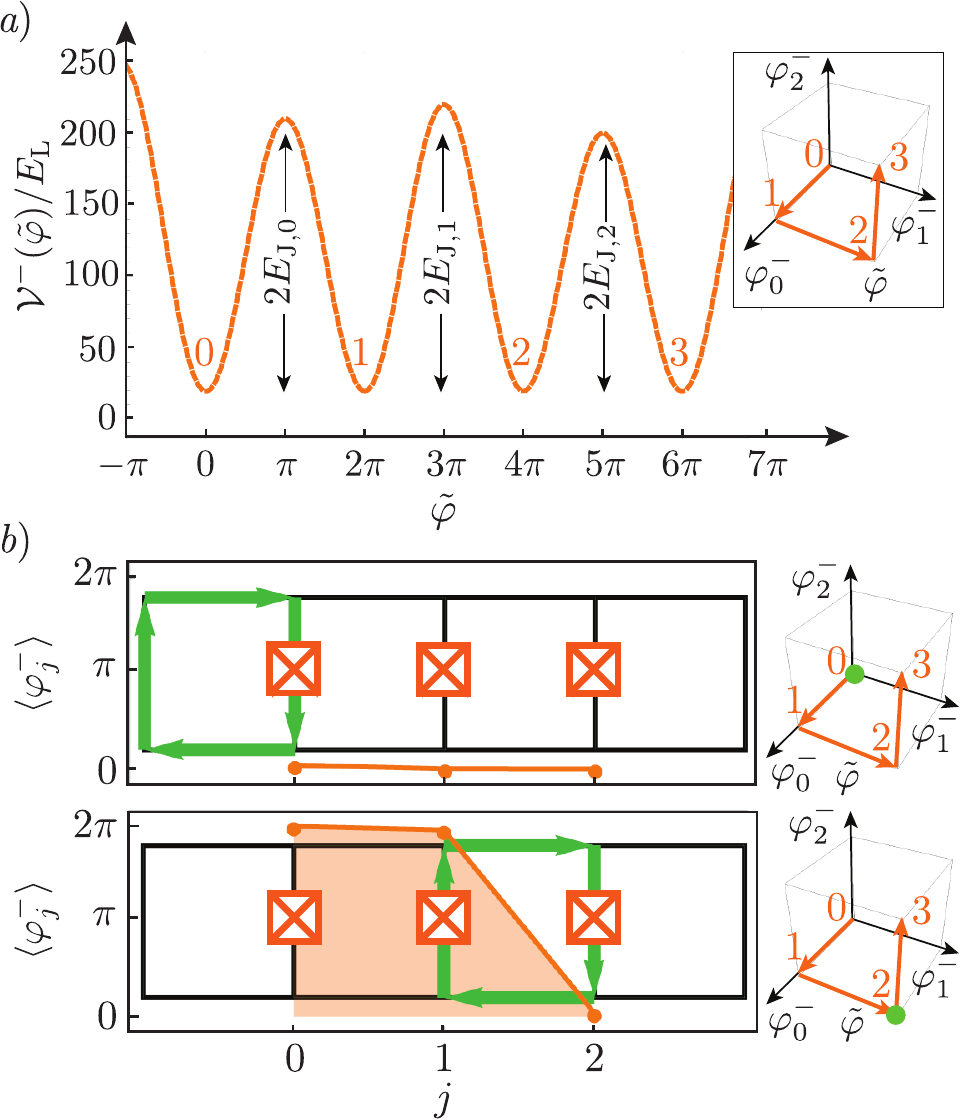}
  \caption{Effective one-dimensional fluxon potential. \textit{a)} The classical potential for 1-fluxon dynamics has four degenerate minima separated by Josephson energy barriers. The insets show that in $( \varphi_0^-,\varphi_1^-,\varphi_2^- )$ space the variable $\tilde{\varphi}$ traverses the edges of a hypercube between the four minima [see Eq.~(\ref{Eq:Kink})]. \textit{b)} The 1-fluxon state consists of a kink in the superconducting phase. If the kink occurs between junctions $j$ and $j+1$, in the limit $E_\text{J} \gg E_{\text{L}}$, the dominant current circulation (green arrows) occurs on the loop delimited by the two junctions. The circulating currents on the neighboring loops are suppressed by a factor $\sim E_\text{J}/E_\text{L}$. \label{Fig:3}}
\end{figure}

In the semiclassical picture, 1-fluxon states correspond to minima of the potential energy $\mathcal{V}^-$ with respect to flux variables $\varphi^-_{i}$, as shown for example in Fig.~\ref{Fig:3}\textit{a}) for $N=3$, describing a single fluxon trapped inside the superinductor ring surrounding the lattice in Fig.~\ref{Fig:2}\textit{b}). Consider the $N+1$ configurations ($k=0,...,N$):
\begin{eqnarray}
  (\varphi_{i}^{-})^{(k)} &\approx& 2\pi ,\text{ for } 0 \leq i < k,
  \nonumber \\
  &\approx& 0,\;\; \text{ for }  k \leq i \leq N. \label{Eq:Kink}
\end{eqnarray}
One-fluxon states correspond to kinks in the expectation value of the field $\varphi^{-}_{i}$ as a function of $i$, as shown in Fig.~\ref{Fig:3}\textit{b}). 

The expressions in Eq.~(\ref{Eq:Kink}) are not exact due to the quadratic contributions of the inductive energy terms $\propto~E_{\text{L}}$. 
These deviations give rise to single vortices of persistent current localized at the position of the kink. The insets of Fig.~\ref{Fig:3}\textit{b}) show expectation values of the currents $\frac{\Phi_0}{2\pi} I_{\text{J},i} = E_{\text{J},i} \sin( \varphi_i^- )$, $\frac{\Phi_0}{2\pi} I_{i}^- 
= \sqrt{2} E_\text{L} (\varphi_{i+1}^- - \varphi_{i}^- - \varphi_{\text{ext},i+1})$. The confinement of the persistent currents is essential to enable the local control of the potential energy, and it follows from the choice of energy scales $E_\text{L}  \ll E_\text{J}$ in Eq.~(\ref{Eq:Vminus}). 

In the 1-fluxon manifold, the relevant variable is the position of the kink. To parametrize this position, we define the variable $\tilde{\varphi}$ along the curve in the $(\varphi_{0}^-,\varphi_{1}^-, ... , \varphi^-_{N-1})$ space which contains the minima of the potential energy, and their connections along classical instanton trajectories \cite{coleman_1988,vainshtein_et_al_1982}. For example, for $N=3$, the potential $\mathcal{V}^-(\tilde{\varphi})$ plotted in Fig.~\ref{Fig:3}a) has degenerate minima at points labeled $0$,...,$3$, corresponding to four classical 1-fluxon states along the curve $\tilde{\varphi}$ represented in the inset. The minima are labeled by the position of the kink, where ``$0$'' stands for no kink, and ``1'' for the kink at the first junction etc. [Fig.~\ref{Fig:3}b)].

The charging energy $\mathcal{T}^-$ gives rise to quantum tunneling between 1-fluxon states. Projecting $\mathcal{H}^-$ into the 1-fluxon manifold yields a quantum tight--binding model
\begin{equation}
  h^- = \sum_{i=0}^{N-1} \epsilon_i |i \rangle \langle i| - \sum_{i=0}^{N-2}  t_{i} | i \rangle \langle i+1| + \text{H.c.}, \label{Eq:TBH}
\end{equation}   
where $|i\rangle$ denotes the 1-fluxon state at $i=0,...,N-1$. We have retained in $h^-$ the next--neighbor contributions only [see Fig.~\ref{Fig:2}\textit{c})], as tunnel rates drop exponentially with distance.  
The on-site energies are $\epsilon_i \approx \frac{1}{2} \hbar \omega_i$ where $\omega_i=\sqrt{8 E^-_{\text{C},i} E_{\text{J},i}}$ is the Josephson plasma frequency. The tunneling rate \cite{matveev_et_al_2002,coleman_1988,vainshtein_et_al_1982, koch_et_al_2007} (the splitting of the $N-$fold degenerate low-lying manifold of classical minima) is exponentially small $t_i \propto e^{- \sqrt{8 E_{\text{J},i}/E^-_{\text{C},i}}}$ and becomes zero in the classical limit $E_{\text{J},i} \gg E^-_{\text{C},i}$. Since the precise value of the numerical prefactor depends on the shape of the potential, in the following we solve for the tunnel rates exactly via numerical diagonalization. 

  \begin{figure}[t!]
    \textit{a})
    \includegraphics[width=0.95\linewidth]{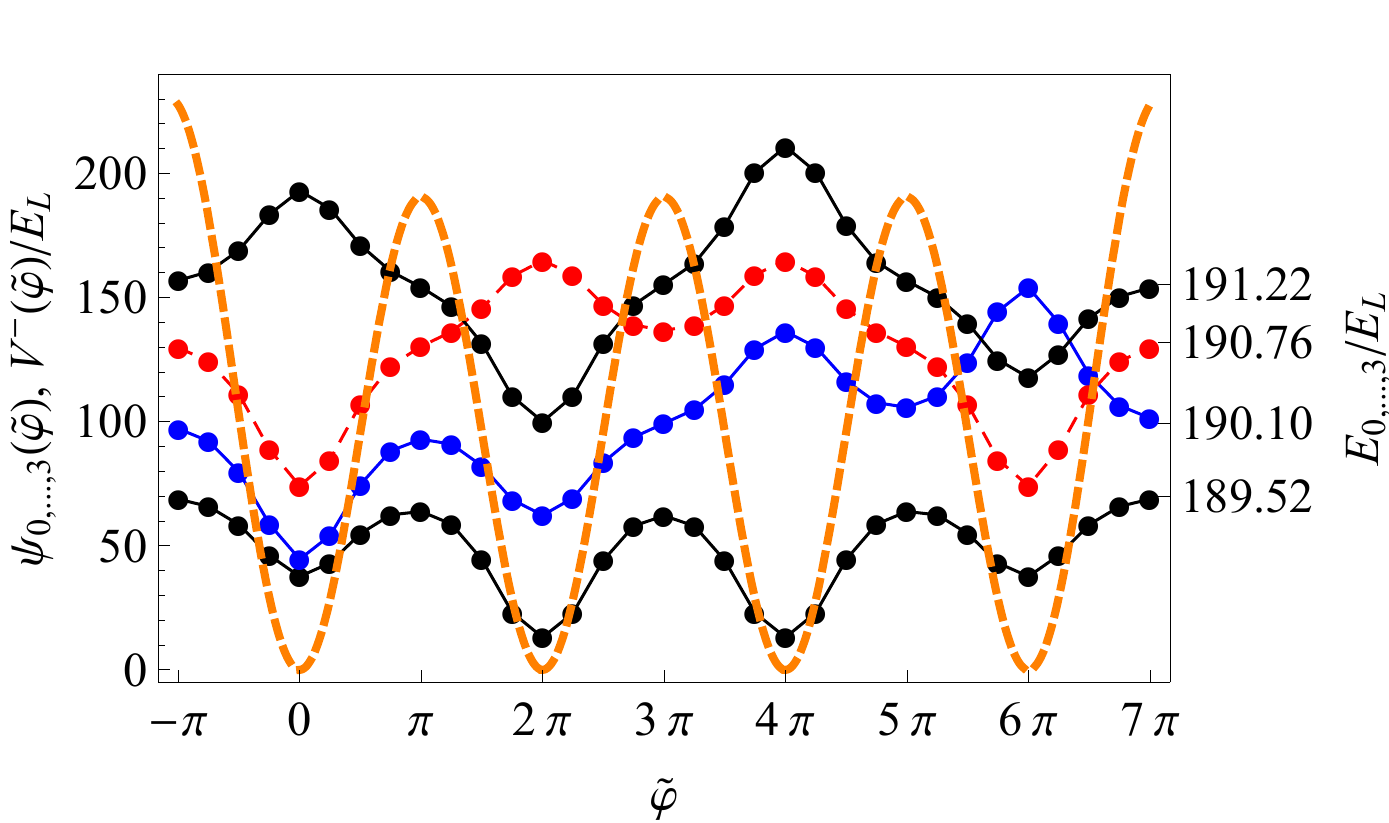}
    \textit{b})
    \includegraphics[width=0.95\linewidth]{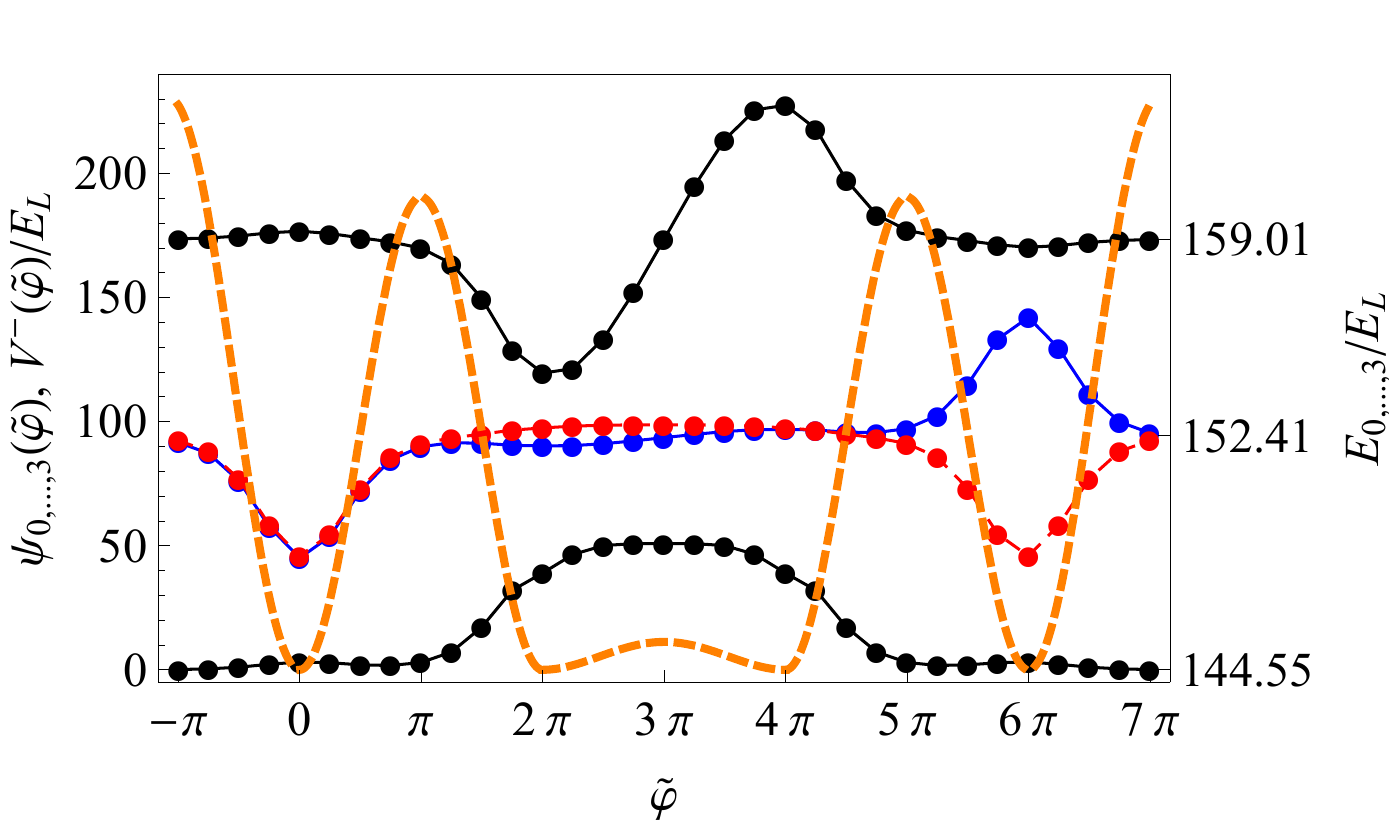}
    \caption{\label{Fig:4} Low-lying spectrum and wavefunctions for the lowest four states (represented ascendingly with respect to their energy, in solid black, dashed red, solid blue, and solid black, respectively) for a three-junction circuit with, $E_\text{C}=10$ and $E_{\text{J}0} = E_{\text{J}2} = 100 = \eta E_{\text{J}1}$ in units of $E_\text{L} = \frac{(\Phi_0/2\pi)^2}{2L}$, with $\eta = 1$ for \textit{a}) and $\eta = 10$ for \textit{b}). We show in orange dashed lines the potential energy $\mathcal{V}^-(\tilde{\varphi})$ (left vertical axis). The wavefunctions in arbitrary units are offset by their eigenenergies (right vertical axis). For \textit{b}), the first and second excited states are intragap boundary-localized excitations of the 1-fluxon tight-binding model. The points represent the values obtained from numerical diagonalization, and the lines are direct connections.} 
  \end{figure}

The low-energy 1-fluxon manifold is separated from the remainder of the spectrum by either a gap of order $(2\pi)^2 E_\text{L}$, corresponding to the creation of fluxon-antifluxon pairs in Eq.~(\ref{Eq:Vminus}), or by an energy scale corresponding to the Josephson plasma frequency. If multiple fluxons are inserted into the array, it is expected that vortex dynamics closely resembles that of a gas of hardcore bosons for energy scales comparable to the bandwidth $t_i$ and far inferior to the gap. In particular, the Mott insulating state of one fluxon per plaquette corresponds to the band insulator obtained by occupying all states of the (band) spectrum of Eq.~(\ref{Eq:TBH}) with $\epsilon_i = \epsilon$ and $t_i = t$. Note that the Hamiltonian for fluxon dynamics $\mathcal{H}^-$ is dual to that of bosons on a two-leg Josephson ladder, which has a rich  ground state phase diagram depending on external flux and boson density \cite{orignac_giamarchi_2001,petrescu_le_hur_2013,piraud_et_al_2014,petrescu_et_al_2017}.

We validate our semiclassical arguments with an exact numerical diagonalization. For this purpose, we consider  $N=3$ junctions
\begin{eqnarray}
  \mathcal{H}^- = 4 E^-_\text{C} \left[(n^-_0)^2 + (n^-_1)^2 + (n^-_2)^2 \right]  + \mathcal{V}^-(\varphi_0^-,\varphi_1^-, \varphi_2^-). \nonumber \\\; \label{EqS:HMinus}
\end{eqnarray}
To numerically diagonalize $\mathcal{H}^-$ we consider the equivalent eigenvalue problem and solve it by a finite-difference method \cite{dempster_et_al_2014} complemented by exact diagonalization [Supplemental Material]. 
We plot the wavefunction  $\psi(\varphi_0^-,\varphi^-_1,\varphi^-_2)$, along the $\tilde{\varphi}$ coordinate, and the eigenvalues of lowest-lying states in Fig.~\ref{Fig:4}\textit{a}).  
Due to the action of the charging (Laplacian) terms, there is some leakage of the wavefunctions along the coordinates perpendicular to the curve parametrized by $\tilde{\varphi}$. This effect is taken into account in the multidimensional numerical diagonalization. 

Tunneling amplitudes can be tuned to yield a topological bandstructure in one dimension. Here, we discuss a fluxon analogue of the Su-Schrieffer-Heeger \cite{su_et_al_1979} tight-binding model, a model originally proposed to describe the electronic structure of polyacetylene, an organic compound that features a Peierls instability, by which consecutive bonds in a one-dimensional tight-binding chain alternate between strong and weak. The Su-Schrieffer-Heeger model sustains a (chiral) symmetry protected topological phase \cite{ryu_hatsugai_2002, wen_2012,bernevig_neupert_2015}. 

The dimerization of the Josephson energy $E_{\text{J}j} = E_\text{J} + (-1)^{j+1} \delta$ with $E_\text{J} > \delta > 0 $ and $j=0,...,N-1$  achieves the Su-Schrieffer-Heeger bandstructure \cite{su_et_al_1979, su_et_al_1980} in the 1-fluxon effective model of Eq.~(\ref{Eq:TBH}). The number $N$ of Josephson junctions for which a pair of topological states is observable is an odd number, with $N \geq 3$. In Fig.~\ref{Fig:4}\textit{b}) we show the low-lying energies and eigenstates for the minimal length $N=3$ and $E_{\text{J}0}=E_{\text{J}2}=10 E_{\text{J}1}$ and $E^-_\text{C} = E_{\text{J}0} / 10$, \textit{i.e.} a three-junction circuit in which the middle junction is more strongly in the quantum regime. The effect of enhanced tunneling on the middle junction is to split the states corresponding to fluxons localized on the two central loops, which leads to a large energy gap. The remaining two intragap states correspond to fluxons localized on the end loops. For increasing system length $N$ the hybridization of the localized end-loop states must vanish exponentially. 

The levels of the dimerized low-energy model, which amount to a gapped conduction band at half-filling \cite{su_et_al_1979}, can be filled as fluxons are added to the system. At half-filling, when one inserts $m=\frac{N+1}{2}+1$ fluxons, the ground state has two intra-gap boundary-localized fluxon excitations \cite{jackiw_rebbi_1976, su_et_al_1980, su_schrieffer_1980, bernevig_neupert_2015}. The boundary states, topologically protected against perturbations in the bulk, could be used for the implementation of a superconducting qubit. This may offer an alternative to fault tolerant quantum computation via topological protection, as embodied for example by the $0-\pi$ qubit \cite{kitaev_1997,doucot_vidal_2002, ioffe_feigelman_2002, kitaev_2006,gladchenko_et_al_2009,brooks_et_al_2013, dempster_et_al_2014}.

In conclusion, we have presented an alternative path to perform quantum simulation, moving away from the well-known microwave photon architectures to a concept based on fluxon dynamics in networks of Josephson junctions. Unlike photons, fluxons can be individually trapped inside superinductor loops, and their number $m$ can be remarkably stable in time, for durations practically infinite compared to the typical experiment timescale. The control and readout of the states could be performed using the standard tools of cQED, while the quantum Hamiltonian of the simulation is encoded in long-lived quantum fluxon states. Dispersive quantum non-demolition measurements \cite{smith_et_al_2016} could be adapted to access the local density of states in such circuits, by using locally coupled RF antennas. These spectroscopic methods go beyond previous direct current transport experiments with Josephson junction networks probing the vortex superfluid and Mott insulating states \cite{van_der_zant_et_al_1992,van_oudenaarden_mooij_1996}. 

We have discussed the possible experimental limitations of this platform and argued that the current quantum fluxon model is robust for networks containing up to order of ten lattice sites, after which the transmission-line modes of the circuit can interfere with the fluxon modes. This limit could be increased by using more sophisticated circuit fabrication technologies, which can remove most of the backplane dielectric via etching, and thus decrease the self capacitance \cite{chu_et_al_2016}. 

The power of the quantum fluxonics concept is illustrated by a circuit implementation of the Su-Schrieffer-Heeger model in the 1-fluxon subspace. Even a relatively simple circuit implementation of this model, with four lattice sites, displays a spectrum including a pair of edge states, which could be used to implement a superconducting qubit.  Finally, we note that beyond the scope of quantum simulation, the concept of quantum fluxonics could be appealing for on-chip quantum state transfer \cite{averin_et_al_2006,deng_averin_2014}, or for quantum signal routing using traveling fluxons \cite{mueller_et_al_2017}.

We are grateful to Richard Brierley, Michel Devoret, Karyn Le~Hur, Boris Malomed, Nick Masluk, Uri Vool and Andrei Vrajitoarea for insightful discussions. AP and HET were supported by the US Department of Energy, Office of Basic Energy Sciences, Division of Materials Sciences and Engineering, under Award No. DE-SC0016011. AVU acknowledges partial support from the Ministry of Education and Science of the Russian Federation in the framework of the contract No.~K2-2016-063. IMP acknowledges funding from the Alexander von Humboldt foundation in the framework of a Sofja Kovalevskaja award endowed by the German Federal Ministry of Education and Research. 

\bibliographystyle{apsrev4-1}
\bibliography{refs}


\begin{widetext}
\clearpage

\begin{center}
\textbf{\large SUPPLEMENTAL MATERIAL:\\ Fluxon-Based Quantum Simulation in \
Circuit QED}
\end{center}
\setcounter{equation}{0}
\setcounter{figure}{0}
\setcounter{table}{0}
\setcounter{page}{1}
\makeatletter
\renewcommand{\theequation}{S\arabic{equation}}
\renewcommand{\thefigure}{S\arabic{figure}}
This Supplemental Material contains a discussion of fluxon insertion inside a superinductor loop, the derivation of the circuit Hamiltonian used in the main text, as well as a detailed description of the numerical methods employed to diagonalize it for a small number of Josephson junctions, $N$.

\section{Fluxon insertion}
In this section, we provide a more detailed discussion of the protocol for fluxon insertion. We consider the circuit in Fig.~\ref{Fig:SILoop}, in which $L$ is a superinductance \cite{masluk_et_al_2012}, as described in the main text, and the loop is closed by an input Josephson junction with Josephson energy $E_{\text{J},\text{in}}$ approximately one hundred times the charging energy $E_{\text{C},\text{in}}$. Before we review the time--dependent protocol introduced in the main text, we derive the equations of motion and the potential energy for the circuit of Fig.~\ref{Fig:SILoop}. The physics of the input junction is analogous to that of a weak link interrupting a loop of superconductor \cite{zimmerman_silver_1965,silver_zimmerman_1967,thouless_1998}. 

We now write a system of classical equations of motion for branch fluxes and currents corresponding to the Josephson junction and the inductor. These can be represented in terms of node variables $\Phi_\text{J} = \phi_1 - \phi_g$ and $\Phi_\text{L} = \phi_1 - \phi_g + \Phi_\text{ext}$, respectively, from which we derive the loop equation for branch fluxes:
\begin{equation}
  \Phi_\text{L} = \Phi_\text{J} + \Phi_\text{ext}. \label{Eq:LoopEq}
\end{equation}
Current conservation at node 1 means
\begin{equation}
  I_\text{J} + I_\text{L} = C_\text{J} \ddot{\phi}_{1}. \label{Eq:NodeEq}
\end{equation}
Equations~(\ref{Eq:LoopEq}) and~(\ref{Eq:NodeEq}) underlie the derivation of the Hamiltonian of the circuit in Fig.~\ref{Fig:SILoop}a) based on the rules of circuit quantization \cite{devoret_1997}. 

The purpose of this section is to derive the potential energy and its stationarity conditions. To this end, let us set the right member of Eq.~(\ref{Eq:NodeEq}) to zero, and denote the loop current symbol $I$, with the following sign convention: 
\begin{equation}
  \label{Eq:Currents}
  I = I_\text{J} = - I_\text{L}.
\end{equation}
The current around the loop can be related to the phase difference across the Josephson junction in the following way. Let 
\begin{equation}
  \varphi_\text{J} \equiv 2\pi \frac{\Phi_J}{\Phi_0} \text{ mod } 2\pi
\end{equation}
be the superconducting phase difference across the Josephson junction. It is useful to explicitly introduce an integer $m$ such that the equality modulo multiples of $2\pi$ becomes
\begin{equation}
  \label{Eq:FluxPhase}
  \varphi_\text{J} = 2\pi \frac{\Phi_J}{\Phi_0} + 2\pi m.
\end{equation}

The phase variable $\varphi_\text{J}$ is defined to be compact on the interval $(-\pi,\pi]$. It is related to the current through the Josephson junction through the Josephson relation
\begin{equation}
  \label{Eq:JosephsonRel}
  I = I_\text{c} \sin\left( \varphi_\text{J} \right), \varphi_\text{J} = \sin^{-1}\left( \frac{I}{I_\text{c}} \right),
\end{equation}
where $I_\text{c}$ is the critical current. It is related to the Josephson energy through the relation $E_{\text{J},\text{in}} = I_\text{c} \Phi_0/(2\pi)$.  

The current $I$ is also related to the flux through the inductor $\Phi_\text{L}$ through the constitutive equation
\begin{equation}
  \label{Eq:PhiL}
  \Phi_\text{L} = - L I,
\end{equation}
where we have used Eq.~(\ref{Eq:Currents}). 

We can now use the Josephson relation~(\ref{Eq:JosephsonRel}), the equation relating the flux and phase variables~(\ref{Eq:FluxPhase}), and the constitutive equation of the inductor~(\ref{Eq:PhiL}) together with the loop equation~(\ref{Eq:LoopEq}) to obtain
\begin{equation}
  - L I = \frac{\varphi_\text{J}}{2\pi} \Phi_0 - m \Phi_0 + \Phi_\text{ext}.
\end{equation}
Rearranging terms, this gives
\begin{equation}
  \label{Eq:FluxoidQuantization}
\frac{\varphi_\text{J}}{2\pi} \Phi_0  + \left( L I   + \Phi_\text{ext} \right ) = m \Phi_0.  
\end{equation}
The quantity on the right-hand side is the London fluxoid. The term in the parentheses is the total flux through the superconducting loop, composed of the kinetic flux $L I$ from the loop inductance $L$ and the external flux $\Phi_\text{ext}$. This is the fluxoid quantization condition \cite{thouless_1998,tinkham_1996}.  

\begin{figure}[t!]
  \textit{a})\includegraphics[width=0.25\linewidth]{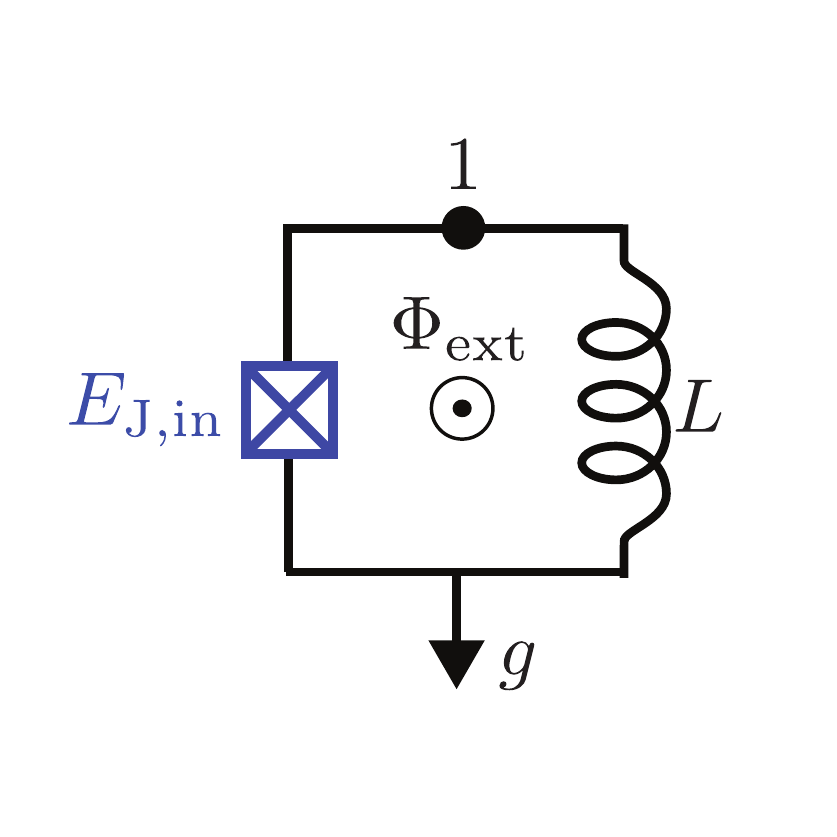}
  \textit{b})\includegraphics[width=0.35\linewidth]{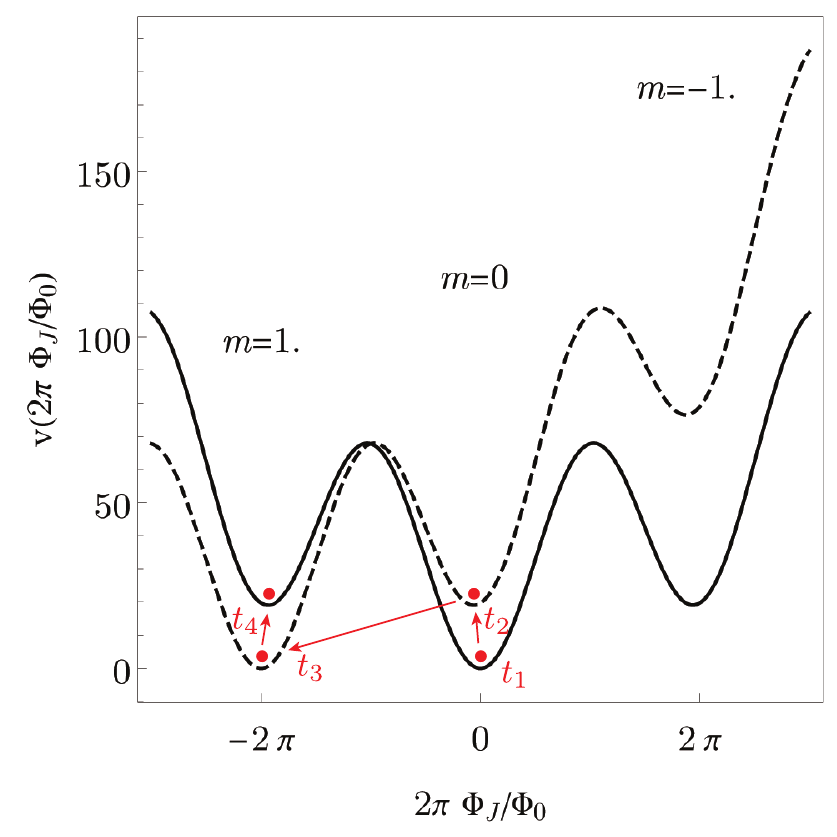}
  \caption{\label{Fig:SILoop} a) Circuit layout illustrating the conventions in the text; b) Reduced potential function $v(\varphi_J)$ for the circuit in a) at $\varphi_\text{ext} = 0$ (black solid) and $\varphi_\text{ext} = 2\pi$ (black dashed). The times $t_1,t_2,t_3,t_4$ correspond to those represented in Figure~1 of the main text, and the red arrows and circle markers indicate the 4 stages of the insertion of 1 fluxon. The integers $m$ above the three central minima indicate the value of the fluxoid, $2\pi m$, from Eq.~(\ref{Eq:FluxoidQuantization}) or its equivalent~(\ref{Eq:varphiJEq}).}
\end{figure}

Using the Josephson relation~(\ref{Eq:JosephsonRel}) in Eq.~(\ref{Eq:FluxoidQuantization}) we arrive at the transcendental equation 
\begin{equation}
  \label{Eq:varphiJEq}
  \varphi_\text{J} - 2 \pi m + 2\pi\frac{\Phi_\text{ext}}{\Phi_0} = - \sin(\varphi_\text{J}).
\end{equation}
Recall that $\varphi_\text{J}$ is defined on the compact interval $(-\pi,\pi]$. Different solutions of the transcendental equation above are obtained by varying $m$ at fixed $\Phi_\text{ext}$. Alternatively, one may use the relation between $\Phi_\text{J}$ and $\varphi_\text{J}$, Eq.~(\ref{Eq:FluxPhase}), and solve a transcendental equation for the real variable, the flux: 
\begin{equation}
  \label{Eq:PhiJEq}
  2\pi \Phi_\text{J}/\Phi_0 + 2\pi \Phi_\text{ext}/\Phi_0 = - \sin(2\pi\Phi_\text{J}/\Phi_0).
\end{equation}
Equations~(\ref{Eq:varphiJEq}) and~(\ref{Eq:PhiJEq}) are equivalent and they serve to distinguish between the compact phase variable $\varphi_\text{J}$ and the real flux variable $\Phi_\text{J}$. The equation for the compact phase variable $\varphi_\text{J}$ necessarily contains the London fluxoid $2\pi m$ [in units of $\Phi_0/(2\pi)$].

Equation~(\ref{Eq:PhiJEq}) is a stationarity condition for the dimensionless potential energy [consistent with the equations of motion~(\ref{Eq:LoopEq}) and~(\ref{Eq:NodeEq})]
\begin{equation}
  v\left(2\pi\frac{\Phi_\text{J}}{\Phi_0}\right) = 
\left[ 1 - \cos \left(2\pi \frac{\Phi_\text{J}}{\Phi_0}\right) \right] + 2\pi \frac{(\Phi_\text{J} + \Phi_\text{ext})^2}{2\Phi_\text{c} \Phi_0},
\end{equation}
where we have introduced the critical kinetic flux $\Phi_\text{c} = L I_\text{c}$. This function is plotted in Fig.~\ref{Fig:SILoop}b) for two values of the external flux $\Phi_\text{ext} = 0$ (solid lines) and $\Phi_0$ (dashed lines). The minima of the potential energy are labeled by their respective values of the fluxoid $2\pi m$, as obtained from the solution to the transcendental equation~(\ref{Eq:varphiJEq}).

The fluxon insertion protocol relies on that of Masluk~\textit{et al} \cite{masluk_et_al_2012}. The input junction is addressable by means of the antenna connected across a shunt capacitance $C_\text{S}$. The superinductor loop is threaded by external flux $\Phi_\text{ext}$. The insertion of one fluxon entails increasing the fluxoid from $m=0$ to $m=1$ in units of the superconducting flux quantum, in the following sequence:  Before $t_1$ at zero external flux, the system is in its classical ground state corresponding to $m=0$. At $t_1$, the flux is increased to $\Phi_0$ maintaining the system in the metastable minimum. Between $t_2$ and $t_3$ a high-amplitude drive is applied to lower the effective Josephson potential $E_\text{J,in}$, which prompts a spontaneous relaxation of the system to the lower energy state at $m=1$. At $t_4$, the flux is turned back to zero, thereby placing the system in an (excited) metastable state at $m=1$. The procedure can be iterated to insert additional fluxons. To insert $m$ fluxons, a field $\Phi_\text{ext} = m \Phi_0$ would be necessary, in order to turn the $m$ fluxon minimum into a global minimum at time $t_2$.

\section{Derivation of the circuit Hamiltonian for the Josephson transmission line}
Consider the circuit in Figure~\ref{Fig:Circuit}. We follow Ref.~[\onlinecite{devoret_1997}] to quantize the circuit. We will generalize our results to $2N$ superconducting islands but keep the calculation concrete at $N=3$ for brevity. Below, $g$ denotes the ground node, to which superconducting island $\alpha, j$, with $\alpha=1,2$ and $j=0,1,2$, is connected via capacitance $C_{\alpha, j}$. The Josephson energy of the $j^{\textit{th}}$ junction is $E_{\text{J},j} = \frac{\hbar I_{\text{c},j}}{2e}$, where $I_{\text{c},j}$ denotes the critical current on the $j^\textit{th}$ junction. The capacitance of each junction is $C_{\text{J},j}$. The minimum spanning tree (MST) covering the 6 active nodes $\alpha j$ for $\alpha=1,2$, $j=0,1,2$ is highlighted in gray in Fig.~\ref{Fig:Circuit}.  The loop equations in terms of branch variables (labeled according to Fig.~\ref{Fig:Circuit}) are:
\begin{figure}[t!]
  \includegraphics[width=0.5 \linewidth]{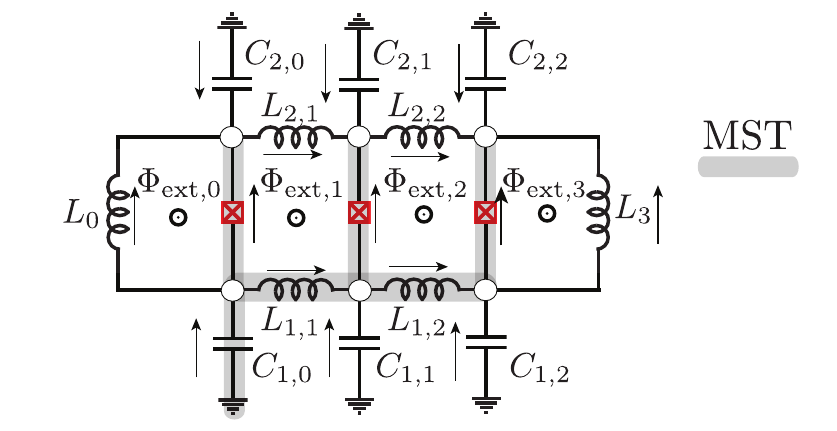}
  \caption{\label{Fig:Circuit} $N=3$ junction circuit with open boundaries. Minimum spanning tree (MST) \cite{devoret_1997} is highlighted in gray. The junctions, not labeled above, are characterized by Josephson energy $E_{\text{J},j}$ and capacitance $C_{\text{J},j}$, $j=0,1,2$.}
\end{figure}

\begin{eqnarray}
  \Phi_{C_{1,1}} - \Phi_{L_{1,1}} - \Phi_{C_{1,0}} &=& 0, 
  \nonumber \\
  \Phi_{C_{1,2}} - \Phi_{L_{1,2}} - \Phi_{L_{1,1}} - \Phi_{C_{1,0}} &=& 0,
  \nonumber \\
  \Phi_{C_{2,0}} - \Phi_{E_{\text{J},0}} - \Phi_{C_{1,0}} &=& 0,
  \nonumber \\
  \Phi_{C_{2,1}} - \Phi_{E_{\text{J},1}} - \Phi_{L_{1,1}} - \Phi_{C_{1,0}} &=& 0,
  \nonumber \\
  \Phi_{C_{2,2}} - \Phi_{E_{\text{J},2}} - \Phi_{L_{1,2}} - \Phi_{L_{1,1}} - \Phi_{C_{1,0}} &=& 0,
  \nonumber \\
  \Phi_{E_{\text{J},0}} - \Phi_{L_0} &=& \Phi_{\text{ext},0},
  \nonumber \\
  \Phi_{E_{\text{J},1}} - \Phi_{L_{2,1}} - \Phi_{E_{\text{J},0}} + \Phi_{L_{1,1}} &=& \Phi_{\text{ext},1},
  \nonumber \\
  \Phi_{E_{\text{J},2}} - \Phi_{L_{2,2}} - \Phi_{E_{\text{J},1}} + \Phi_{L_{1,2}} &=& \Phi_{\text{ext},2}, \nonumber \\
  -\Phi_{E_{\text{J},2}} + \Phi_{L_3} &=& \Phi_{\text{ext},3}.
 \label{Eq:loop}
\end{eqnarray}

The branch fluxes for branches that belong to the MST can be reexpressed in terms of node fluxes, 
\begin{eqnarray}
 \Phi_{E_{\text{J},0}} = \phi_{2,0} - \phi_{1,0},&\;& \Phi_{E_{\text{J},1}}=\phi_{2,1}-\phi_{1,1}, \nonumber \\
 \Phi_{E_{\text{J},2}} = \phi_{2,2} - \phi_{1,2},&\;& \Phi_{L_{1,1}} = \phi_{1,1} - \phi_{1,0}, \nonumber \\
 \Phi_{L_{1,2}} = \phi_{1,2} - \phi_{1,1},&\;& \Phi_{C_{1,0}} = \phi_{1,0} - \phi_g. \label{Eq:Branches1}
\end{eqnarray}

Replacing these into the loop Eqs.~(\ref{Eq:loop}) we obtain
\begin{eqnarray} \label{Eq:Branches2}
  \Phi_{C_{1,1}} = \phi_{1,1} - \phi_g, \; \Phi_{C_{1,2}} = \phi_{1,2} - \phi_g, \;\;\;\;\;\;\;\; \\
  \Phi_{C_{2,0}} = \phi_{2,0} - \phi_g, \; \Phi_{C_{2,1}} = \phi_{2,1} - \phi_g, \nonumber \\
  \Phi_{C_{2,2}} = \phi_{2,2} - \phi_{g}, \nonumber \\
  \Phi_{L_0} = \phi_{2,0} - \phi_{1,0} - \Phi_{\text{ext},0}, \; \Phi_{L_3} = \phi_{2,2} - \phi_{1,2} + \Phi_{\text{ext},3},\; \nonumber \\
  \Phi_{L_{2,1}} = \phi_{2,1} - \phi_{2,0} - \Phi_{\text{ext},1}, \; \Phi_{L_{2,2}} = \phi_{2,2} - \phi_{2,1} - \Phi_{\text{ext},2}. \nonumber
\end{eqnarray}

Substituting (\ref{Eq:Branches1}) and (\ref{Eq:Branches2}) into Kirchoff node equations, we find equations of motion 
\begin{eqnarray}
  \alpha, i&& \nonumber \\
  1,0&:&\; -\frac{\phi_{2,0} - \phi_{1,0} - \Phi_{\text{ext},0}}{L_0} - I_{\text{c},0} \sin\left(2\pi \frac{\phi_{2,0}-\phi_{1,0}}{\phi_0} \right) + \frac{\phi_{1,1}-\phi_{1,0}}{L_{1,1}} = C_{1,0} (\ddot{\phi_{1,0}} - \ddot{\phi_g}) - C_0 (\ddot{\phi_{2,0}} - \ddot{\phi_{1,0}}), \nonumber \\
  2,0&:&\; + \frac{\phi_{2,0} - \phi_{1,0} - \Phi_{\text{ext},0}}{L_0} + I_{\text{c},0} \sin\left(2\pi \frac{\phi_{2,0}-\phi_{1,0}}{\phi_0} \right) - \frac{\phi_{2,1} - \phi_{2,0} -\Phi_{\text{ext},1}}{L_{2,1}} = C_{2,0} (\ddot{\phi_{2,0}} - \ddot{\phi_g}) + C_0 (\ddot{\phi_{2,0}} - \ddot{\phi_{1,0}}), \nonumber \\
  1,1&:&\; \frac{\phi_{1,1}-\phi_{1,0}}{L_{1,1}} - I_{\text{c},1} \sin\left( 2\pi \frac{\phi_{2,1}-\phi_{1,1}}{\phi_0}\right) - \frac{\phi_{1,2}-\phi_{1,1}}{L_{1,2}} = C_{1,1} (\ddot{\phi_{1,1}}-\ddot{\phi_g}) - C_1(\ddot{\phi_{2,1}} - \ddot{\phi_{1,1}}), \\
  2,1&:& \frac{\phi_{2,1} - \phi_{2,0} - \Phi_{\text{ext}1}}{L_{2,1}} + I_{\text{c},1} \sin\left( 2\pi \frac{\phi_{2,1}-\phi_{1,1}}{\phi_0}\right) - \frac{\phi_{2,2} - \phi_{2,1} - \Phi_{\text{ext}2}}{L_{2,2}} = C_{2,1} (\ddot{\phi_{2,1}} - \ddot{\phi_g}) + C_1(\ddot{\phi_{2,1}}-\ddot{\phi_{1,1}}), \nonumber \\
  1,2&:& + \frac{\phi_{1,2} - \phi_{1,1}}{L_{1,2}} - I_{\text{c},2} \sin\left( 2\pi \frac{\phi_{2,2} - \phi_{1,2}}{\phi_0} \right) - \frac{\phi_{2,2} - \phi_{1,2} + \Phi_{\text{ext}3}}{L_3} = C_{1,2} (\ddot{\phi_{1,2}} - \ddot{\phi_g}) - C_2 (\ddot{\phi_{2,2}}- \ddot{\phi_{1,2}}), \nonumber \\
  2,2&:& + \frac{\phi_{2,2} - \phi_{2,1} - \Phi_{\text{ext},2}}{L_{2,2}} + I_{\text{c},2} \sin\left( 2\pi \frac{\phi_{2,2} - \phi_{1,2}}{\phi_0}\right) + \frac{\phi_{2,2}-\phi_{1,2} + \Phi_{\text{ext},3}}{L_3} = C_{2,2}( \ddot{\phi_{2,2}} - \ddot{\phi_g}) + C_2 (\ddot{\phi_{2,2}}- \ddot{\phi_{1,2}}). \nonumber 
\end{eqnarray}
These are Euler--Lagrange equations for the following Lagrangian (expressed now in terms of $N$; to retrieve the previous equations, one would set $N=3$):
\begin{eqnarray}
  \label{Eq:MathcalL}
  \mathcal{L} &=& \sum_{j=0}^{N-1} \frac{1}{2} C_j (\dot{\phi_{2,j}}-\dot{\phi_{1,j}})^2 + \sum_{j=0}^{N-1}\sum_{\alpha=1,2} \frac{1}{2} C_{\alpha i} (\dot{\phi_{\alpha i}}- \dot{\phi_g})^2 -\sum_{j=0}^{N-2} \left[ \frac{(\phi_{1,j+1}-\phi_{1,j})^2}{2 L_{1,j+1}} + \frac{(\phi_{2,j+1}-\phi_{2,j} - \Phi_{\text{ext},j+1})^2}{2 L_{2,j+1}}  \right] \nonumber \\  
  &&\;\;\;\;\;\;\;\; -\frac{(\phi_{2,0} - \phi_{1,0} - \Phi_{\text{ext},0})^2}{2L_0} - \frac{(\phi_{2,N-1} - \phi_{1,N-1} + \Phi_{\text{ext},N})^2}{2L_N} + \sum_{j=0}^{N-1} E_{\text{J},j} \left[1- \cos\left( 2\pi \frac{\phi_{2,j}-\phi_{1,j}}{\phi_0}\right) \right].
\end{eqnarray}

Now set the longitudinal inductances to be all equal, $L_{\alpha, i} = L$, and the terminal inductors to a value that ensures that all loop inductances are constant across the circuit $L_0 = L_N = 2 L$. Further let the capacitance to ground of each superconducting island be $C_{\alpha, i} = C_0$, for $i=0,...,N-1$ and $\alpha=1,2$. These assignments agree with the particular choices denoted in Fig.~2\textit{b}) in the main text. We now introduce new coordinates 
\begin{equation}
  \label{Eq:AandSVars}
  \phi_{j}^{\pm} = \phi_{2,j} \pm \phi_{1,j}. 
\end{equation}
In terms of these fields the charging energy is rearranged into
\begin{eqnarray}
\frac{1}{2} C_0 (\dot{\phi_{j,0}} - \dot{\phi_{g}})^2 + \frac{1}{2} C_0( \dot{\phi_{j,1}}-\dot{\phi_g})^2  \equiv \frac{1}{2} C_0 (A^2 + B^2 ) = \frac{1}{2} C_0 \frac{(A+B)^2 + (A-B)^2}{2} = C_0 \left[ (\dot{\phi_j^+}/2 - \dot{\phi_g})^2 + (\dot{\phi_j^-}/2)^2  \right] \nonumber \\
\end{eqnarray}
and the longitudinal inductive elements give rise to:
\begin{eqnarray}
  \frac{(\phi_{1,j+1} - \phi_{1,j})^2}{2L} + \frac{(\phi_{2,j+1}- \phi_{2,j}-\Phi_{\text{ext},j+1})^2}{2L} = \frac{1}{4L} \left[ \left( \phi_{j+1}^+ - \phi_j^+ - \Phi_{\text{ext},j+1}  \right)^2 + \left( \phi_{j+1}^- - \phi_j^- - \Phi_{\text{ext},j+1} \right)^2 \right].
\end{eqnarray}
Additionally, the inductive terms for the two end loops transform to
\begin{eqnarray}
  \frac{(\phi_{2,0}-\phi_{1,0}-\Phi_{\text{ext},0})^2}{2\times 2 L} = \frac{(\phi_0^- - \Phi_{\text{ext},0})^2}{2\times 2 L},\;\;    \frac{(\phi_{2,N-1}-\phi_{1,N-1}+\Phi_{\text{ext},N})^2}{2 \times 2 L} = \frac{(\phi_{N-1}^- + \Phi_{\text{ext},N})^2}{2 \times 2 L}.
\end{eqnarray}

\begin{figure}[ht!]
  \textit{a})\includegraphics[width=0.95\linewidth]{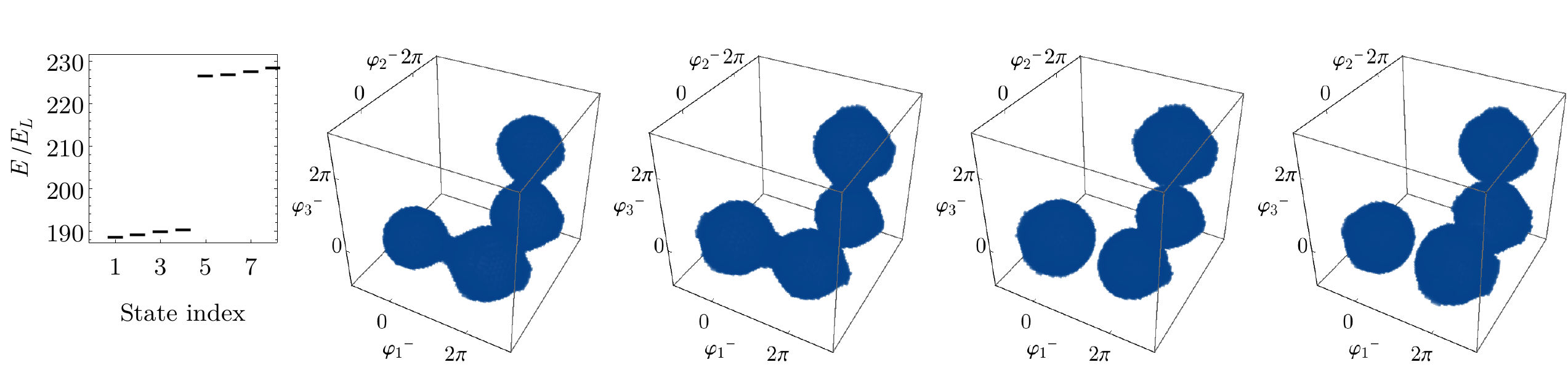}
  \textit{b})\includegraphics[width=0.95\linewidth]{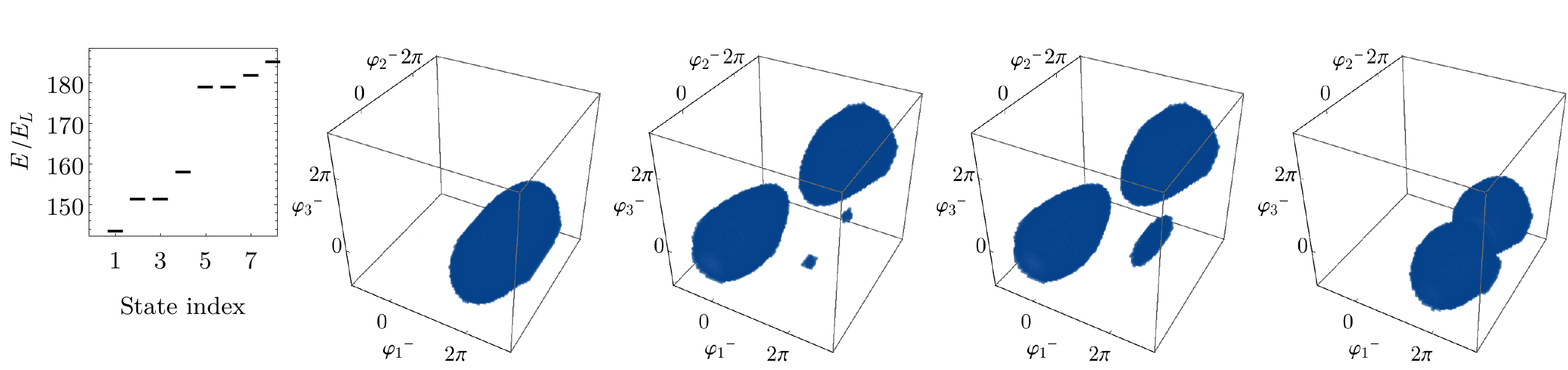}
  \caption{\label{FigS:2}Three-dimensional density plots (opaque volumes signify that the absolute value of the probability density exceeds $10^{-3}$) for the first four eigenstates of $\mathcal{H}^-$ in Eq.~(\ref{EqS:HMinus}) obtained from the finite-differences solution with $N_p = 17$. The low-lying energies are represented in the leftmost panels. We set $E_\text{C} = E_{\text{J},0} \times 10^{-1}$, $E_{\text{J},0}=\eta \times E_{\text{J},1} = E_{\text{J},2}$ in units of $E_\text{L} = \frac{(\Phi_0/2\pi)^2}{2L}$, with \textit{a}) $\eta = 1$  and \textit{b}) $\eta = 10$  corresponding to the values chosen in Fig.~4 of the main text. In \textit{a}), the eigenvalues of the first five states, in units of $E_\text{L}$, are $189.51, 190.10, 190.75, 191.22,$ and $227.49$, respectively. In \textit{b)}, the first five eigenvalues are $144.55, 152.34, 152.41, 159.01$ and $179.91$, respectively.}
\end{figure}

In terms of the new coordinates introduced in~(\ref{Eq:AandSVars}) the Lagrangian of Eq.~(\ref{Eq:MathcalL}) becomes
\begin{eqnarray}
  \mathcal{L} = \sum_{j=0}^{N-1} \frac{C_{\text{J},j}}{2} (\dot{\phi_j^-})^2 + \sum_{j=0}^{N-1} \frac{C_{0}}{4} \left[ (\dot{\phi_j^+} - 2\dot{\phi_g})^2 + (\dot{\phi_j^-})^2 \right] - \sum_{j=0}^{N-2} \frac{1}{4 L} \left[ \left( \phi_{j+1}^+  - \phi_j^+  - \Phi_{\text{ext},j+1} \right)^2 + \left( \phi_{j+1}^- - \phi_j^- - \Phi_{\text{ext},j+1} \right)^2 \right] \nonumber \\
  - \frac{1}{4L} \left[ \left( \phi_0^- - \Phi_{\text{ext},0} \right)^2 + \left( \phi_{N-1}^- + \Phi_{\text{ext},N} \right)^2 \right] + \sum_{j=0}^{N-1}  E_{\text{J},j} \left[1-\cos\left( 2\pi \frac{\phi_j^-}{\Phi_0} \right)\right]. \;\;\;\;\;\;\;\;
\end{eqnarray}
The canonically conjugate momenta corresponding to the variables introduced in Eq.~(\ref{Eq:AandSVars}) are
\begin{eqnarray}
  \frac{\partial \mathcal{L}}{\partial \dot{\phi_j^-}} \equiv Q_j^- = (C_{\text{J},j} + C_{0}/2) \dot{\phi_j^-},\; 
  \frac{\partial \mathcal{L}}{\partial \dot{\phi_j^+}} \equiv Q_j^+ = (C_{0}/2) (\dot{\phi_j^+} - 2 \dot{\phi_g}), \;
  \frac{\partial \mathcal{L}}{\partial \dot{\phi_g}} \equiv Q_g = -\sum_{j=0}^{N-1} C_{0} (\dot{\phi_j^+} - 2 \dot{\phi_g}).
\end{eqnarray}
After a Legendre transform, $\mathcal{H} \equiv Q_g \dot{\phi_g} + \sum_{j=0}^{N-1} \sum_{\alpha=\pm} Q_j^\alpha \dot{\phi_j^\alpha} - \mathcal{L}$, and promoting classical degrees of freedom to quantum operators, we find
\begin{eqnarray}
  \mathcal{H} &=& \sum_{j=0}^{N-1} \frac{(Q_j^+)^2}{2(C_{0}/2)} + \sum_{j=0}^{N-1} \frac{(Q_j^-)^2}{2(C_{\text{J},j} + C_{0}/2)} + \sum_{j=0}^{N-2} \frac{1}{4 L} \left[ \left( \phi_{j+1}^+ - \phi_j^+ -\Phi_{\text{ext},j+1} \right)^2 + \left( \phi_{j+1}^- - \phi_j^- - \Phi_{\text{ext},j+1} \right)^2 \right] \nonumber \\
  &&+ \frac{1}{4L} \left[ \left( \phi_0^- - \Phi_{\text{ext},0} \right)^2 + \left( \phi_{N-1}^- + \Phi_{\text{ext},N} \right)^2 \right] - \sum_{j=0}^{N-1} E_{\text{J},j} \left[1-\cos\left( 2\pi \frac{\phi_j^-}{\phi_0} \right)\right].
\end{eqnarray}
We introduce, as in the main text, a dimensionless variable for the flux $\varphi_{j}^{\alpha} = 2\pi \phi_{j}^{\alpha}/\Phi_0$ and the canonically conjugate Cooper pair number $n_{j}^{\alpha} = \frac{Q_{j}^{\alpha}}{2e}$ for $j=0,...,N-1$ and $\alpha = \pm$. We also introduce energy scales associated with charging and inductive circuit elements 
\begin{equation}
  E_{\text{C}}^+ = \frac{e^2}{2(C_0/2)},\; E_\text{L} = \frac{\left[\Phi_0/(2\pi)\right]^2}{2L},\; E^-_{\text{C},j} = \frac{e^2}{2(C_{\text{J},j}+C_0/2)},
\end{equation}
as well as dimensionless flux variables
\begin{equation}
  \varphi_{\text{ext},j} = \frac{2\pi}{\Phi_0} \Phi_{\text{ext},j}.
\end{equation}
The Hamiltonian reads 
\begin{equation}
  \mathcal{H} = \mathcal{H}^+ + \mathcal{H}^-,
\end{equation}
where
\begin{eqnarray}
  \mathcal{H} &=& \sum_{j=0}^{N-1} 4 E_{\text{C}}^+ (n_j^+)^2 + \sum_{j=0}^{N-1} 4 E_{\text{C},j}^- (n_j^-)^2 + \sum_{j=0}^{N-2} \frac{E_\text{L}}{2} \left[ \left( \varphi_{j+1}^+ - \varphi_j^+ -\varphi_\text{ext,j+1} \right)^2 + \left( \varphi_{j+1}^- - \varphi_j^- - \varphi_\text{ext,j+1} \right)^2 \right] \nonumber \\
  &&+ \frac{E_\text{L}}{2}\left[ \left( \varphi_0^- - \varphi_{\text{ext},0} \right)^2 + \left( \varphi_{N-1}^- + \varphi_{\text{ext},N} \right)^2 \right] - \sum_{j=0}^{N-1} E_{\text{J},j} \left[1-\cos\left( \varphi_j^- \right)\right].   \label{Eq:FinHam}
\end{eqnarray}
This is the Hamiltonian used in the main text.

\section{Numerical methods}
In this section we detail the solution to Eq.~(13) of the main text:
\begin{equation}
   \mathcal{H}^- \approx 4 E_\text{C} \left[(n^-_1)^2 + (n^-_2)^2 + (n^-_3)^2 \right]  + \mathcal{V}^-(\varphi_0^-,\varphi_1^-, \varphi_2^-), \label{EqS:HMinus}
\end{equation}
where one flux quantum is threaded through the entire circuit. The latter condition makes the classical global minimum correspond to fluxoid $m = 1$ [in analogy to the point marked $t_3$ in Fig.~\ref{Fig:SILoop}b)]. We choose a gauge such that $\varphi_{\text{ext},0}= 2\pi$ and $\varphi_{\text{ext},i}=0$ for $i=1,2,3$.  Moreover, making the inductances of the 4 elementary loops in the circuit equal ensures that the global minimum of the potential energy is four-fold degenerate -- this is the underlying tight-binding lattice.

Writing $n_i^- = - i \frac{\partial}{\partial \varphi_i^-}$ the associated Schr\"odinger equation takes the form of a differential eigenvalue equation 
\begin{equation}
  \mathcal{H}^-\left(\left\{ \left. -i \frac{\partial}{\partial \varphi_i^-}, \varphi_i^-  \right\vert i = 0,..,2 \right\}\right) \psi = E \,\psi
\end{equation}
This eigenvalue equation can be solved by finite-difference methods \cite{dempster_et_al_2014}. With one flux quantum threaded through the loop, as explained in the previous paragraph, the lowest energy manifold will only contain one-fluxon states, and therefore we only consider the interval $(\varphi_0^-,\varphi_1^-, \varphi_2^-) \in [-\pi,3\pi]\times[-\pi,3\pi]\times[-\pi,3\pi]$. This interval symmetrically contains the minima at $0$ and $2\pi$. We cover this interval by a uniform mesh of $N_p$ points in each of the three directions. Local minima of the classical potential outside of the first octant are higher than the ones inside it by an energy approximately equal to $(2\pi)^2 E_\text{L}$, as follows from the expression of the potential energy in Eq.~(\ref{Eq:FinHam}), and their influence is neglected. We adapt the mesh size so that in the classical limit, corresponding to vanishing charging energies $E_\text{C} = 0$, the lowest energy eigenvalues and the corresponding wavefunctions agree with the minima of the classical potential. In Fig.~\ref{FigS:2} we show results for the uniform and dimerized lattices for a computation corresponding to $N=3$ junctions and mesh size $N_p=17$ along each axis. Diagonalization was performed with a Jacobi-Davidson routine in the Mathematica Package. 

\end{widetext}
\end{document}